\documentclass[12pt]{article}

\usepackage{scicite}
\usepackage{graphicx}

\usepackage{times}

\newenvironment{sciabstract}{%
\begin{quote} \bf}
{\end{quote}}

\newcounter{lastnote}
\newenvironment{scilastnote}{%
\setcounter{lastnote}{\value{enumiv}}%
\addtocounter{lastnote}{+1}%
\begin{list}%
{\arabic{lastnote}.}
{\setlength{\leftmargin}{.22in}}
{\setlength{\labelsep}{.5em}}}
{\end{list}}

\title{Phosphorus in the Young Supernova Remnant Cassiopeia A}

\author
{Bon-Chul Koo$^{1\ast}$, Yong-Hyun Lee$^{1}$, Dae-Sik Moon$^{2,3,4}$, Sung-Chul Yoon$^{1}$\\
 \& John C. Raymond$^{5}$\\
\\
\normalsize{$^{1}$Department of Physics and Astronomy, Seoul National University,}\\
\normalsize{Seoul 151-747, Korea}\\
\normalsize{$^{2}$Department of Astronomy and Astrophysics, University of Toronto,}\\
\normalsize{Toronto, Ontario M5S 3H4, Canada}\\
\normalsize{$^{3}$Space Radiation Laboratory, California Institute of Technology,}\\
\normalsize{Pasadena, CA 91125, USA}\\
\normalsize{$^{4}$Visiting Brain Pool Scholar, Korea Astronomy and Space Science Institute,}\\
\normalsize{Daejeon 305-348, Korea}\\
\normalsize{$^{5}$Harvard-Smithsonian Center for Astrophysics,}\\
\normalsize{60 Garden Street, Cambridge, MA 02138, USA}
}

\date{}

\begin{document}


\baselineskip24pt

\def\farcsec{\mbox{$.\!\!^{\prime\prime}$}}%
\def\simlt{\lower.5ex\hbox{$\; \buildrel < \over \sim \;$}}
\def\simgt{\lower.5ex\hbox{$\; \buildrel > \over \sim \;$}}
\def\ea{et al.\ }
\def\kms{km s$^{-1}$}
\def\msun{{$M_\odot$}}
\def\lsun{{L_\odot}}
\def\nH{{n_{\rm H}}}
\def\mh{{M_{\rm H}}}
\def\xpfe{$X({\rm P}/{\rm Fe})$}
\def\fp{F_{\rm [P II]1.189}}
\def\ffe{F_{\rm [Fe II]1.257}}
%

\maketitle


\begin{sciabstract}

Phosphorus ($^{31}$P), which is essential for life, is
thought to be synthesized in massive 
stars and dispersed into interstellar space when
these stars explode as supernovae (SNe). 
Here we report on near-infrared spectroscopic
observations of the
young SN remnant Cassiopeia A, which show that
the abundance ratio of phosphorus to the major nucleosynthetic
product iron ($^{56}$Fe) in
SN material is up to 100 times the average ratio of the 
Milky Way, confirming that 
phosphorus is produced in SNe.
The observed range is compatible with predictions from SN nucleosynthetic
models but not with the scenario in which 
the chemical elements in 
the inner SN layers are completely mixed by hydrodynamic 
instabilities during the explosion. 

\end{sciabstract}


Phosphorus (P) is an indispensable ingredient 
for life together with carbon, hydrogen, nitrogen, oxygen, and sulfur (S). 
In our solar system, 
its abundance relative to hydrogen is
$2.8\times 10^{-7}$ by number, which is 50 to 1900 times less 
than those of 
the other life-keeping $\alpha$ elements\cite{asplund2009}.
The abundance of P in the diffuse interstellar medium and 
stars in our galaxy's disk is  
comparable with that of the solar 
system or the cosmic abundance, with some dependence 
on metallicity\cite{lebouteiller2005,caffau2011}.
P is believed to be mainly formed in massive 
[$\simgt 8~M_\odot$] stars by neutron capture on silicon (Si) 
in hydrostatic neon-burning shells in the pre-SN stage and
also in explosive carbon and neon burning layers during SN explosion
\cite{woosley1995,noteSNIa}.

Freshly synthesized P should thus be found in 
young core-collapse SN remnants (SNRs) resulting from the 
explosion of massive stars.
Cassiopeia A (Cas A) 
is the youngest confirmed core-collapse SNR in our galaxy;
it has been extensively studied in all wavebands (Fig. 1). 
It is located at a distance of 3.4 kpc\cite{reed1995} and thought to 
be the remnant of a SN event in C.E. $1681\pm19$ \cite{fesen2006}.
The spectrum of the light echo from the SN event indicated a SN of Type IIb that 
had a small hydrogen envelope at the time of explosion\cite{krause2008}. 
The total estimated mass of the SN ejecta is 2 to 4~\msun, 
and the inferred initial mass of the progenitor star ranges 
from 15 to 25~\msun\cite{hwang2012,young2006}.
Emission from P II was previously detected from Cas A\cite{gerardy2001},
but no analysis of the emission line has been done.

We conducted near-infrared (NIR) spectroscopic observations
of the main SN ejecta shell using the TripleSpec spectrograph
mounted on the Palomar 5-m Hale telescope in 2008 (Fig. 1) 
\cite{noteSM}.
TripleSpec provides
continuous wavelength coverage from 0.94 to 2.46 $\mu$m at
medium spectral resolving power of $\sim 2,700$.
By analyzing the spectra, we identified 
63 knots of emission and, for each knot, measured the radial velocities and  
fluxes of emission lines, including [P II] lines\cite{noteSM}. 

The knots show
distinct spectroscopic and kinematic properties depending on
their origins (Fig. 2).
The knots with strong [S II] lines (Fig. 2, red symbols) have high 
($\simgt 100$~\kms) radial velocities, and most of them have  
strong [P II] lines too, suggesting that the production of P is 
tightly correlated with the production of S. 
Some knots without apparent [S II] lines  
have low radial velocities ($\simlt 100$~\kms) and strong 
He I lines (FIg. 2, green symbols).
Their properties match those of 
``quasi-stationary flocculi", the material lost from the 
hydrogen envelope of the progenitor during its 
red-supergiant phase as slow wind\cite{vandenbergh1971,lee2013}.
The rest of the knots without [S II] lines (Fig. 2, blue symbols) 
have little He I emission, and most of them have high radial velocities. 
They have 
[Fe II] lines as strong as those from the other knots. 
These knots are probably pure iron (Fe) material from complete Si-burning 
in the innermost region of the SN, corresponding to 
the pure Fe ejecta detected in x-rays\cite{hwang2012}. 
They are found mainly in the southern bright [Fe II] filament, where 
the x-ray--emitting Fe ejecta is not prominent (Fig. 1).

The P/Fe abundance ratio of the knots 
can be derived from the flux ratio of [P II] 1.189 $\mu$m and 
[Fe II] 1.257 $\mu$m lines ($\fp/\ffe$).
These two lines have comparable excitation 
energies and critical densities, which together 
with the fact that the ionization energies of P and Fe 
are comparable (10.49 and 7.90 eV) greatly simplifies
the abundance analysis\cite{oliva2001}.
(This is not the case for [S II] lines that have considerably
higher excitation energies.) 
We assume that the line-emitting region is at $T_e=10,000$~K and has an uniform
density, with equal fractions of singly ionized ions.
This simple model yields P/Fe abundances 
accurate to within $\sim 30$\% mostly 
for both SN and circumstellar knots\cite{noteSM}.
The observed $\fp/\ffe$ ratios (Fig. 3) imply that 
the P-to-Fe abundance ratio by number, \xpfe, of the
SN ejecta knots is up to 100
times higher than the cosmic abundance 
$X_\odot({\rm P}/{\rm Fe})=8.1\times 10^{-3}$ \cite{asplund2009},
whereas the knots of the circumstellar
medium have ratios close to the cosmic abundance.
The enhanced P abundance over Fe in these SN ejecta knots
is direct evidence for the in situ production of P in Cas A.

The observed range of \xpfe\ is compatible with the 
local nucleosynthetic yield of P in SN models.
The internal
chemical composition and the resulting \xpfe\
profile of a spherically symmetric SN model for
progenitor mass of 15 M$_\odot$ is shown in Fig. 4 \cite{rauscher2002}.
In the oxygen-rich layer, P 
is enhanced, whereas Fe is slightly depleted because of neutron capture.
($^{56}$Fe and $^{58}$Fe are the two major Fe isotopes 
in the oxygen-rich layer.)
This makes \xpfe\ higher by two orders of magnitude 
than that of the 
outer He-rich layer, which is essentially equal to the 
cosmic abundance. 
The extended P enhancement in 
$M/M_{\rm core}=0.55$ to 0.73 is the result of
hydrostatic burning in pre-SN, whereas
the bump at $M/M_{\rm core}\sim 0.5$
is due to explosive burning during the explosion.
The knots that we observed are probably from
somewhere in the oxygen-rich layer. 
It is, however, difficult
to precisely determine the knots' original positions in the progenitor
star from the comparison of
the measured P/Fe ratio with the model prediction because
the local \xpfe\ are very much model-dependent (Fig. 4B) and subject to
further modifications owing to the Rayleigh-Taylor instability
during the explosion\cite{kifonidis2003}.
We instead assumed that the knots have been expanding
freely, with an expansion rate of 
$2800$~km s$^{-1}$ pc$^{-1}$\cite{delaney2010}.
We then converted the obtained space velocities to mass coordinates
using the ejecta velocity profile of a 
15~\msun\ SN IIb model\cite{woosley1994}
scaled to the chemical structure of the model in Fig. 4\cite{rauscher2002}.
The resulting mass coordinates have considerable uncertainties 
because of model-dependency but
provide rough locations of the knots (Fig. 4B).

\xpfe\ of the [P II] line--emitting ejecta knots
fall into the \xpfe\ range of the oxygen-rich layer in the model,
whereas the \xpfe\ of the ``pure''
Fe knots are often less than the cosmic abundance (Fig. 4B).
The fact that the circumstellar knots have \xpfe\ close to 
the cosmic 
abundance--$X({\rm P}/{\rm Fe})\sim 2 X_\odot({\rm P}/{\rm Fe})$--gives 
confidence in the derived \xpfe\ in the core. 
The extended \xpfe\ range over nearly two orders of magnitude
may be explained by the hydrodynamic chemical mixing during 
the SN explosion.
But, our result  does
not support a complete mixing of the core below the He-rich layer
because in such a case, \xpfe\ will be much lower as marked by the brown arrow
in Fig. 4B; the available yields in the literature imply  
\xpfe=0.01 to 0.05 for a SN of 15 to 25 \msun\
\cite{woosley1995,rauscher2002,kobayashi2006,chieffi2013},
which is represented by the solid part of the arrow, 
or 0.01 to 0.15 when the unusually high P yield of the 
20~\msun\ model of \cite{rauscher2002} is included.
The detection of P-depleted, pure Fe material
probably produced in the innermost, complete Si-burning layer
also indicates that these dense SN
ejecta materials largely retain their original abundance.

The high \xpfe\ ratio, in principle, could also result 
from the depletion of Fe in the gas phase if 
Fe atoms are locked in dust grains.
In Cas A, a substantial 
amount ($\sim 0.1$~\msun) of newly formed 
dust grains in the SN ejecta has been indeed 
detected\cite{rho2008,sibthorpe2010,barlow2010}.
But, they are most likely silicate dust,  
with only 1\% of Fe in the form of FeS\cite{nozawa2010}.
Also, the fact that [P II] lines in the 
SN ejecta are much stronger than  
in the circumstellar knots (Fig. 2) 
is best explained by the 
enhanced abundance of P rather than by the depletion of Fe 
onto dust.

\bibliography{scibib}
\bibliographystyle{Science}

{}


\begin{scilastnote}

 \item[\bf Acknowledgements:] 
This work was supported by Basic Science Research (NRF-2011-0007223) 
and International Cooperation in Science abd Technology 
(NRF-2010-616-C00020) programs 
through the National Research Foundation of Korea (NRF) funded by the Ministry of 
Education, Science and Technology, 
and also by the Korean Federation of Science and Technology Societies (KOFST). 
D.-S. M. acknowledges support from the the Natural
Science and Engineering Research Council of Canada. 
We thank M. Muno for his help in observations and 
J.-J. Lee for providing the {\it Chandra} 1Ms x-ray images. 

\end{scilastnote}

\noindent
{\bf Supplementary Materials} \\
www.sciencemag.org/cgi/content/full/science.1243823/DC1 \\
Materials and Methods \\
Figs. S1 to S4 \\
Tables S1 to S2 \\ 
References (29--49)

\clearpage

\begin{figure}[t]
\begin{center}
 \includegraphics[width=5.5in]{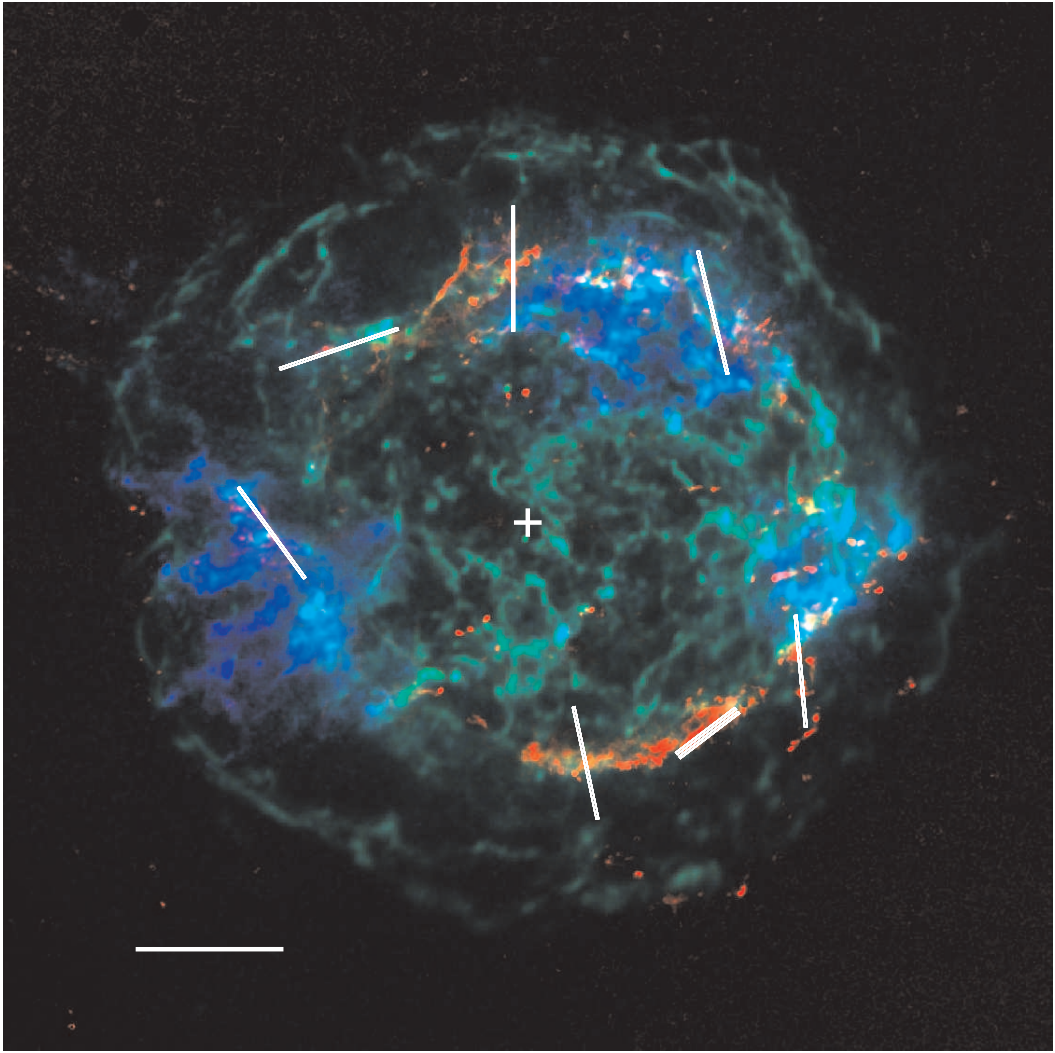}
 \caption{
{\bf Fig. 1. Three-color composite image of the young supernova remnant Cas A.}
Red is a Palomar [Fe II] 1.644-$\mu$m image representing 
SN material at $\sim 10^4$~K,
green is a {\it Chandra} x-ray continuum (4.20 to 6.40 keV) image representing
hot gas and relativistic particles that are heated by SNR blast wave,
and blue is a {\it Chandra} Fe K (6.52 to 6.94 keV) image 
representing SN material at $\sim 2\times 10^7$~K, respectively.
The slit positions in our NIR spectroscopic observations 
are marked by thin white bars. 
The central white plus symbol represents the position of the SN explosion center 
determined from proper motions\cite{thorstensen2001}.
The scale bar in the 
lower left represents an angular scale of $1'$, which corresponds to 
1 pc at the distance (3.4 kpc) of the SNR.}
\label{fig1}
\end{center}
\end{figure}

\begin{figure}[t]
\begin{center}
 \includegraphics[width=6.0in]{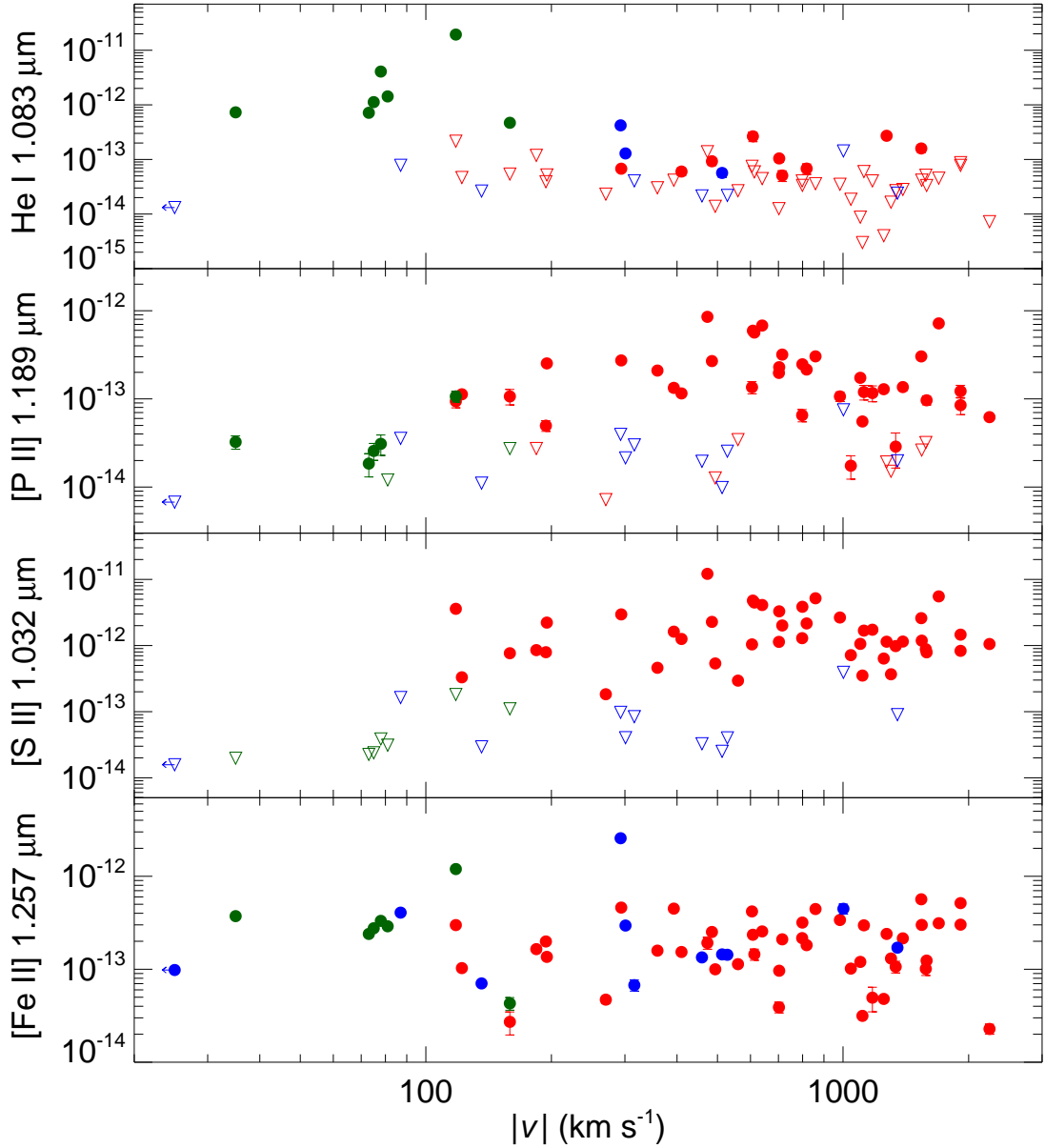}
 \caption{
{\bf Fig. 2. NIR line fluxes  
of the Cas A's knots as a function of their 
radial velocities.} NIR line fluxes are in erg per square centimeter per second. 
Shown are He I 1.083 $\mu$m ($^3P\rightarrow {^3S_1}$), 
[P II] 1.189 $\mu$m ($^{1}D_{2}\rightarrow {^{3}P_{2}}$), 
[S II] 1.032 $\mu$m (${^{2}P_{3/2}}\rightarrow {^{2}D_{5/2}}$), 
and [Fe II] 1.257 $\mu$m ($a{^{4}D_{7/2}}\rightarrow a{^{6}D_{9/2}}$) lines 
from top to bottom. The knots not detected in 
the line emission are marked by open, upside-down triangles
representing $3\sigma$ upper limits. 
The fluxes are corrected 
for extinction. The different colors represent knots of different 
characteristics: red symbols, knots with strong [S II] lines; green symbols, 
knots without [S II] lines but with strong He I lines at low radial velocities 
($\simlt 100$ km s$^{-1}$); blue symbols, knots without [S II] lines and 
with little He I emission 
}
\label{fig2}
\end{center}
\end{figure}

\begin{figure}[t]
\begin{center}
 \includegraphics[width=5.5in]{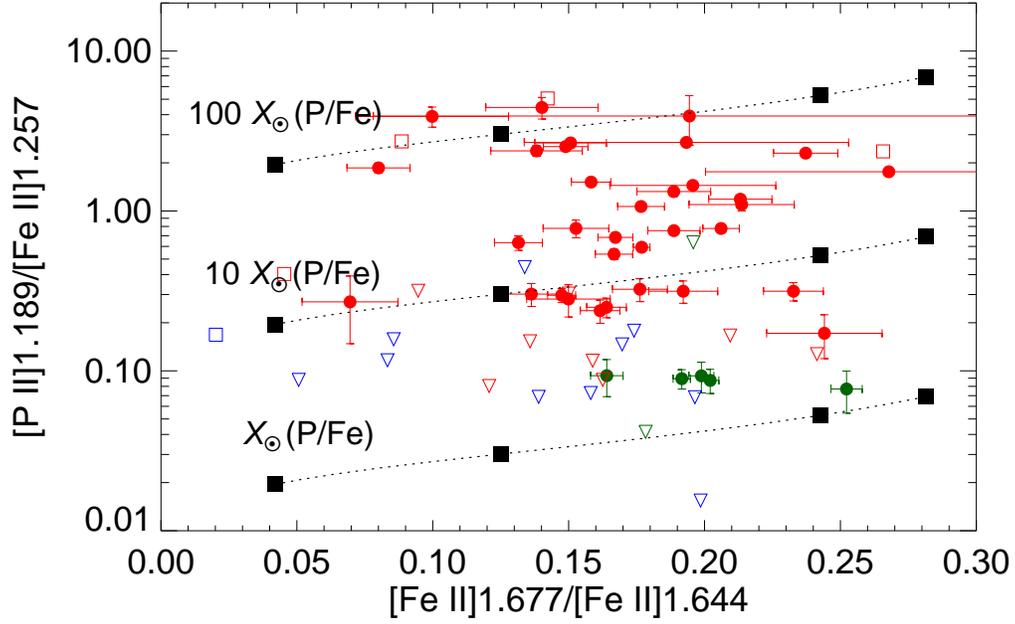}
 \caption{
{\bf Fig. 3. [P II] 1.189/[Fe II] 1.257 versus density-sensitive 
[Fe II] line ratios of the Cas A's knots.} 
The knots are marked in the same colors as in Fig. 2. 
The open, upside-down triangles represent the knots 
undetected in the [P II] line emission, whereas the open squares 
represent the knots undetected in 
[Fe II] 1.677 $\mu$m line emission including the one 
undetected in either line emission at (0.020, 0.17).
The dotted lines show theoretical line ratios for 
\xpfe=1, 10, and 100 times the cosmic abundance. 
The solid squares along the lines represent the ratios at 
$n_e= 10^3, 10^4, 10^5$, and $10^6$~cm$^{-3}$ from left to right. 
The [Fe II] line ratio in the abscissa is 
essentially given by a function of $n_e$, and 
the observed ratios indicate that 
$n_e =3\times 10^3$ to $2\times 10^5$ cm$^{-3}$. 
}
\label{fig3}
\end{center}
\end{figure}

\begin{figure}[t]
\begin{center}
 \includegraphics[width=5.8in]{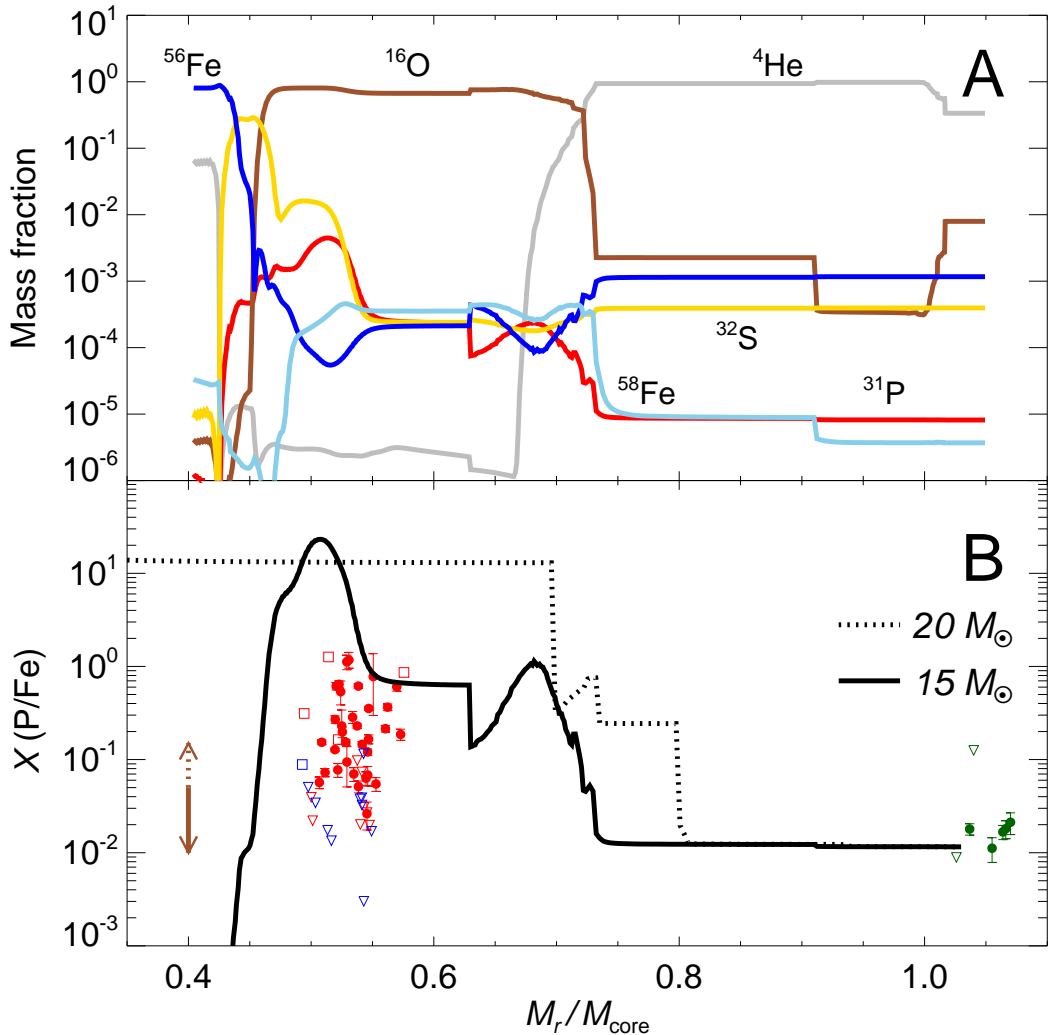}
 \caption{
{\bf Fig. 4. Comparison of the abundance 
ratio \xpfe\ of the Cas A's knots with predictions from SN nucleosynthetic 
models.} 
({\bf A}) Internal, 
post-SN chemical composition of a 15~\msun\ SN 
model \cite{rauscher2002}. 
The abscissa is 
the mass inside a radius $r$ from the explosion center 
normalized by the mass of the SN core which is
4.163~\msun. The mass inside the ``mass cut''
(1.683~\msun\ or $M_r/M_{\rm core}=0.40$)   
becomes a stellar remnant, whereas the rest is ejected 
with the explosion. Most of the mass outside the core would 
have been lost as a stellar wind for SN Type IIb such as Cas A.
In a real SN, the profile
will be smoother because of the hydrodynamic mixing of material.
({\bf B}) Profile of the abundance ratio \xpfe\ 
of the 15~\msun\ SN (solid line).  
The Fe abundance includes all stable Fe isotopes.
The observed \xpfe\ ratios of the Cas A SN ejecta knots 
are marked in the same colors and symbols as in Fig. 3.  
The knots of circumstellar medium (green symbols) are 
marked at $M_r/M_{\rm core}>1$. 
For comparison, the dotted line shows
\xpfe\ profile of a 20~\msun\ SN\cite{rauscher2002} with a 
very high P yield because of unusual convection
mixing the Si and oxygen layers before the explosion. 
The other typical SN models of 15 to 25 \msun\ 
in the literature  
have chemical structures similar to that of the 15 \msun\ SN model here. 
The brown arrow to the left represents the expected range of \xpfe\ 
for a complete mixing of the core below the He-rich layer.
}
\label{fig4}
\end{center}
\end{figure}

\end{document}


\title{}

\def\farcsec{\mbox{$.\!\!^{\prime\prime}$}}%
\def\simlt{\lower.5ex\hbox{$\; \buildrel < \over \sim \;$}}
\def\simgt{\lower.5ex\hbox{$\; \buildrel > \over \sim \;$}}
\def\ea{et al.\ }
\def\hb{\hfill\break}
\def\xpfe{$X({\rm P}/{\rm Fe})$}
\def\xsfe{$x({\rm S})/x({\rm Fe})$}
\def\xps{$x({\rm P})/x({\rm S})$}
\def\xsfep{$x({\rm S}^+)/x({\rm Fe}^+)$}
\def\fion{f_{\rm ion}}
\def\fionp{f_{\rm P II}}
\def\fions{f_{\rm S II}}
\def\fionfe{f_{\rm Fe II}}
\def\np{N({\rm P})}
\def\ns{N({\rm S})}
\def\nfe{N({\rm Fe})}
\def\npp{N({\rm P}^+)}
\def\nsp{N({\rm S}^+)}
\def\nfep{N({\rm Fe}^+)}
\def\pp{${\rm P}^+$}
\def\sp{${\rm S}^+$}
\def\fep{${\rm Fe}^+$}
\def\fp{F_{\rm [P II]}}
\def\fs{F_{\rm [S II]}}
\def\ffe{F_{\rm [Fe II]}}
\def\ppline{[P II] 1.189}
\def\spline{[S II] 1.032}
\def\fepline{[Fe II] 1.257}
\def\fep{${\rm Fe}^+$}
\def\mum{$\mu$m}
\def\ncr{$n_{\rm cr}$}
\def\nion{N_{\rm ion}}
\def\kms{km s$^{-1}$}

\begin{center}
{\large\bf Supplementary Materials} \\
\end{center}

\noindent
{\large\bf S1. Observation and data reduction}

The near-infrared (NIR) spectroscopic observations of 
Cassiopeia A (Cas A) were carried out with the TripleSpec spectrograph 
mounted on the 5-m Palomar Hale telescope on June 29 and August 8, 2008. 
TripleSpec is a slit-based NIR cross-dispersion echelle
spectrograph covering the entire NIR atmospheric windows 
simultaneously with a spectral resolving
power $R\approx2500$--3000 \cite{wilson2004,herter2008}.
The slit width and length of TripleSpec are $1''$ and $30''$, 
respectively.
We obtained spectra at eight slit positions 
well spread over the bright, main shell of Cas A
(see Fig. 1 in the Report).
At each position, reference background spectra were obtained either 
within the slit, by observing the target at the two nod positions 
at $1/3$ (A) and $2/3$ (B) locations along the slit 
in an ABBA pattern, e.g., see \cite{houck2004},
or outside the slit if the field was too crowded to obtain useful background
spectra within the slit. The background spectra were then subtracted out.
In Fig. 1, the positions represented by long ($45''$) white bars 
are where the spectra were obtained in the ABBA pattern, while 
the positions represented by the $30''$-long white bars 
are where the reference spectra were obtained outside the slit.
The total 
exposure time per slit ranged from 300 sec to 1,800 sec. 

For the data reduction, we followed the standard procedure 
of pre-processing 
including dark subtraction, flat-fielding correction, and bad pixel removal.
We then obtained wavelength solutions using the OH airglow
emission lines in the band,
and corrected the wavelength to the heliocentric reference frame.
The sky background including OH lines
were subtracted out as much as possible
using the dithered or background frames. 
The flux calibration was done by comparing the spectra of the nearby 
A0V standard star (HD 223386) observed
just before or after the target observations to the Kuruz model spectrum.
We estimate a systematic uncertainty of 30\% in the absolute flux,
but this  uncertainty did
not enter into the analysis of relative abundances.  

\begin{figure}[t]
\begin{center}
 \includegraphics[width=7.0in]{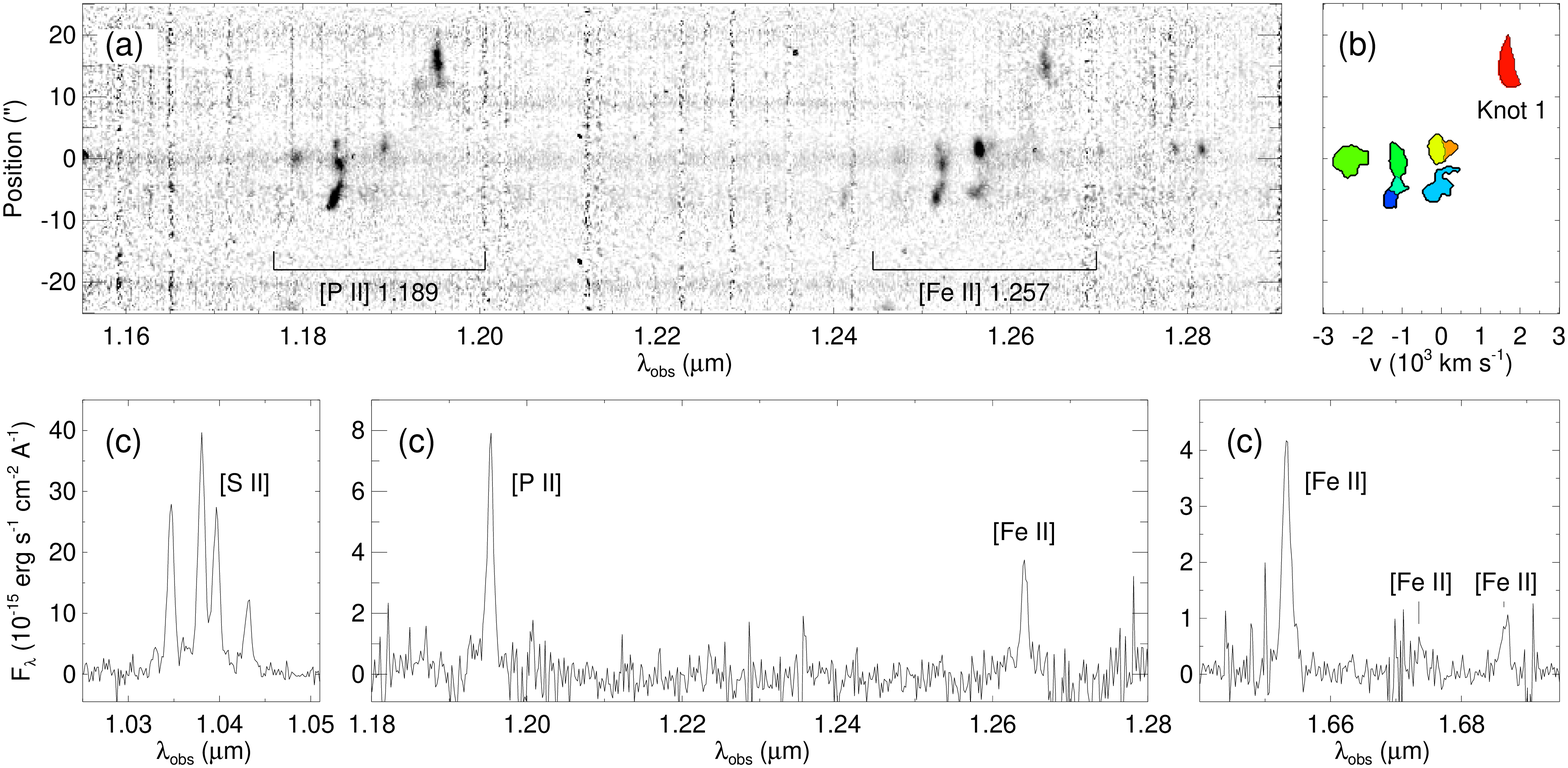}
 \caption{
\small (a) Portion of the two-dimensional dispersed image of Slit 1 
showing [P II] 1.189 $\mu$m and [Fe II] 1.257 $\mu$m lines.
(b) Color-coded `mask image' showing 
the areas in the two-dimensional dispersed image occupied by individual 
knots identified from the {\it Clumpfind} routine\cite{williams1994}. 
The abscissa is the radial velocity determined from the Doppler shifts 
of the lines.
(c) One dimensional spectrum of Knot 1 extracted from (a) using the mask
in (b).
}
\end{center}
\end{figure}

Along a slit, there are usually several [Fe II]-line emitting 
filaments or knots of distinct kinetic properties, 
so that, in a two-dimensional dispersed image, we observe 
complex emission features distributed along both the space and 
velocity (or wavelength) dimensions. 
For example, Fig. S1 (a) shows a portion of the processed 
dispersed image of Slit 1, where 
we see more than five distinct ``knots'' emitting both 
[P II] 1.189 and [Fe II] 1.257 $\mu$m lines.
In order to study the physical conditions of the emitting regions,
we first need to separate out individual emission features. 
We used the clump-finding algorithm 
{\it Clumpfind}\cite{williams1994}, which is developed for the 
identification of clumps in molecular clouds.   
Using a set of contour levels provided as an input parameter,
the routine searches local maxima, decomposes clumps, 
and determines their locations and sizes.
We used a set of contour levels
starting at 2$\sigma$ of the background RMS noise 
with a 2$\sigma$ increment. 
The identification has been made in the dispersed image of the [Fe II] 
1.644 $\mu$m line which is essentially identical to that of the 
[Fe II] 1.257 $\mu$m line in Fig. S1(a). If the [Fe II] line is weaker 
than the [S III] 0.953 $\mu$m or [P II] 1.189 $\mu$m lines, 
we used the latter. 
Fig. S1 (b) shows the result of applying {\it Clumpfind}
to Slit 1, which yielded eight knots in total. 
These ``masks'' in dispersed images 
are then used to extract 
one-dimensional spectra of the knots.
Fig.~S1 (c) shows the extracted one-dimensional spectrum of Knot 1.
The line fluxes are derived from Gaussian least-squares fittings of 
the spectra. 

The derived line parameters of 
the five emission lines used in our analysis are listed in Table~S1.
The radial velocities are those of [Fe II] 1.257 $\mu$m lines, which 
agree with those of the other lines within  
about 100 km s$^{-1}$, comparable with the spectral resolution of our data.
Note that the line fluxes are extinction corrected
based on the ratio of
[Fe II] 1.257 and 1.644 $\mu$m lines.
Since these two lines share the same upper level ($a^{4}D_{7/2}$) 
(see the next section),
their line ratio is determined by the ratio of $A_{21}\nu_{21}$ where
$A_{21}$ is Einstein $A$ coefficients for spontaneous transition from
the upper (2) to the lower (1) state and
$\nu_{21}$ is the frequency of the line. 
We adopt 
the recent theoretical result of \cite{deb2011}, which gives  
the intrinsic ratio 
$F_{1.257}/F_{1.644}=1.36$. The deviation of the observed line ratio 
from this is due to the extinction, and the corresponding 
corrections are applied to the other lines using the extinction curve of the
carbonaceous-silicate grain model with Milky Way size distribution for
$R_{V}=3.1$ \cite{weingartner2001}. 

We note that there is a weak [Fe II] line at 1.18848 $\mu$m
($a^{2}G_{7/2}\rightarrow a^{4}D_{7/2}$)  
that overlaps with [P II] 1.189 (1.18861) $\mu$m line.
Its expected intensity relative to the  
[Fe II] 1.257 $\mu$m line is $\simlt 10^{-5}$ 
at the electron densities $\simlt 10^5$~cm$^{-3}$, 
and consequently its contribution to 
the [P II] line flux should be negligible. 
The identification of this line reported in previous 
studies ({\it 11, 35})
is most likely a misidentification of the [P II] line. 

%
\begin{figure}[t]
\begin{center}
\includegraphics[width=5.5in]{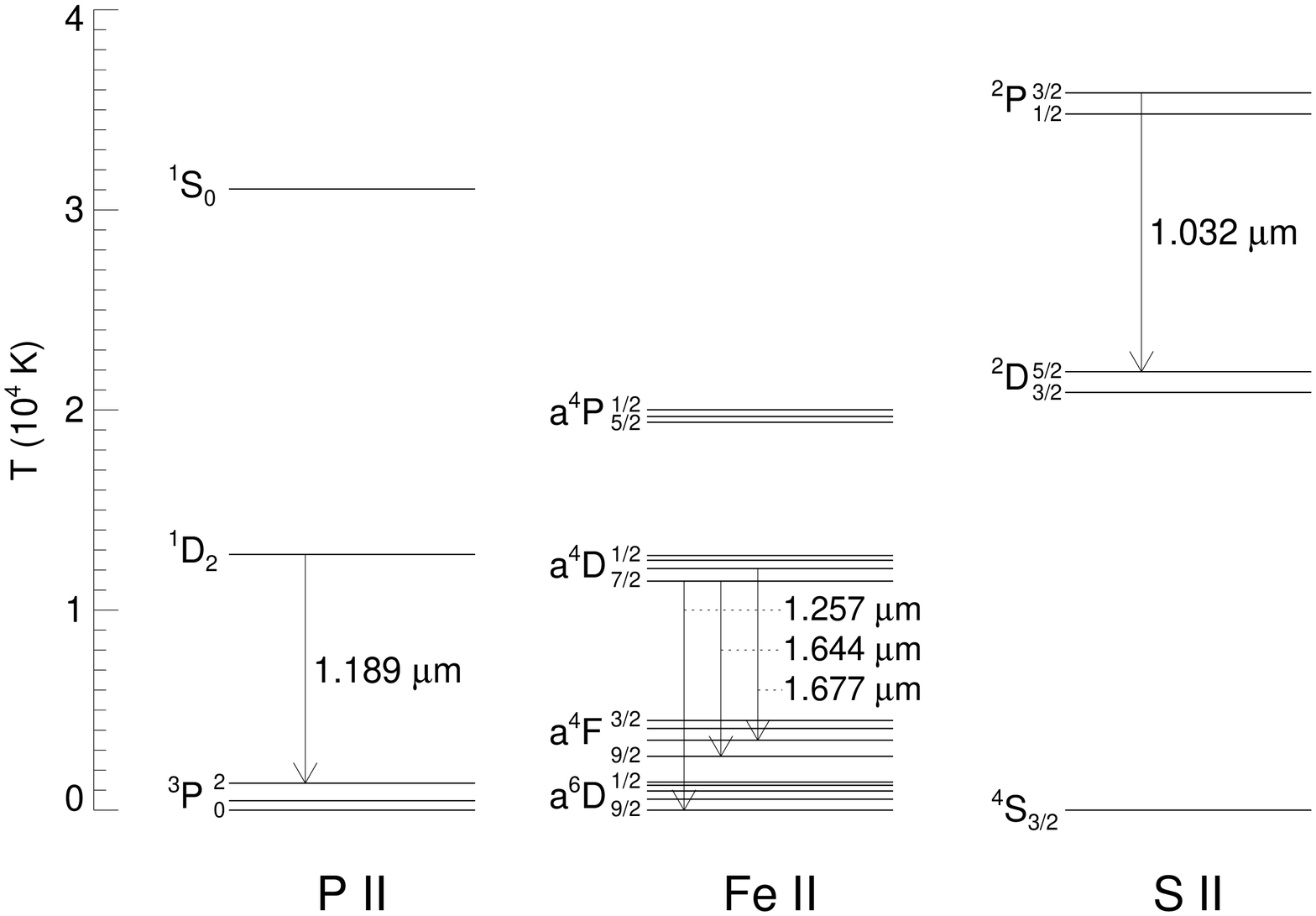}
\caption{
Energy-level diagrams for lowest terms of P II, Fe II, and S II in their ground 
electronic configurations. The emission lines used in this work are indicated
by arrows with their wavelengths (in $\mu$m) labeled. 
The energies of 
upper terms are indicated on the left in temperature scale. 
The separations 
of the fine structure levels are magnified for clarity.  
}
\end{center}
\end{figure}

\bigskip
\noindent
{\large\bf S2. Abundance analysis of [P II] and [Fe II] emission lines}

In this section, we describe the basic properties of the 
forbidden lines that we use in our analysis 
and how we analyze them to derive the abundance ratios. 

\noindent
{\bf Near-infrared forbidden lines of P$^+$ and Fe$^{+}$}

The lines that we use in our analysis are the forbidden lines from P II and Fe II.
The energy level diagrams of the ground terms of these ions are shown in Fig. S2.
P II is a p$^2$-ion and has three low-lying terms. The ground term ($^{3}P$) 
is split into three fine-structure levels, and the transition from
$^{1}D_{2}$ to these levels produces three lines in the NIR band.
Their wavelengths are 1.18861, 1.14713, and 1.12583 \mum, and,
using the Einstein coefficients in Table S2, their intensity ratios are  
$1:0.380:1.37\times 10^{-4}$. 
We use the dominant 1.18861 \mum\ line in our analysis.
Fe II has four ground terms, each of which has
closely-spaced 3--5 levels to form a 16 level system. 
The transitions among these levels produce many lines 
that appear in the near- to mid-infrared bands\cite{pradhan2011, koo2013}. 
The lines to be analyzed in this work are the  
1.25702, 1.64400, and 1.67734 $\mu$m lines. The first two lines
are the two strongest [Fe II] lines, and as we pointed out 
in section S1, their intrinsic ratio is 
constant (=1.36).

In  Fig. S2, we also show the energy level diagram of S II 
for completeness. The transitions among the fine 
structure levels $^2P_{3/2,1/2}$ and $^2D_{5/2,3/2}$ 
produce four adjacent forbidden lines,
the strongest of which is the 1.032 $\mu$m line
(see Fig. S1).
Note that the excitation energy of the \spline\ \mum\ line 
is significantly higher than those of [P II] and [Fe II] 
lines, which makes the ratio of \spline\ $\mu$m line 
to the other lines sensitive to $T_e$.  
In contrast, the excitation energies of the upper states of \ppline\ and \fepline\ 
\mum\ lines are 
comparable to each other (Table S2), 
and we can derive a reliable abundance ratio using these two lines. 

\noindent
{\bf Formulation of abundance analysis}

The general expression for the flux 
(erg cm$^{-2}$ s$^{-1}$) of an optically thin, 
ionic spectral line from a knot 
is given by the integration 
of surface brightness ($I_{21}$) over the solid angle of the source:
\begin{equation}
F_{21}=\int I_{21} d\Omega  = {h\nu_{21}\over 4\pi d^2} N_2 A_{21}  
   = {h\nu_{21}\over 4\pi d^2} A_{21} f_2 f_{\rm ion} N_Z 
\end{equation}
where $N_2$ is the number of ions in the upper state,
$N_Z$ the total number of element Z, 
$\fion(\equiv N_{\rm ion}/N_{\rm Z})$ the fractional 
ionization of the element, 
$f_2(\equiv N_2/N_{\rm ion})$ 
the fractional population of the upper level, and $d$ the distance 
to the source.
This, together with the line parameters in Table S2, 
gives the flux ratio of the 
[P II] 1.189 \mum\ line to the [Fe II] 1.257 \mum\ line 
\begin{equation}
{F_{\rm [P II]1.189}\over F_{\rm [Fe II]1.257}} 
= 2.87 { f_{\rm ^1D_2, PII} \over f_{\rm a^{4}D_{7/2}, Fe II} } 
{\fionp\over\fionfe} X({\rm P/Fe}),
\end{equation}
where $X({\rm P/Fe})\equiv N_{\rm P}/N_{\rm Fe}$ is 
the abundance ratio that we intend to obtain.

The fractional population of the upper level ($f_2$) can be calculated  
for given $n_e$ and $T_e$ by solving rate equations. 
When the density is low, collisional excitation and de-excitation are 
negligible compared to spontaneous radiative transition, and vice 
versa. For level $i$, the critical density can be defined by
\cite{draine2011}
\begin{equation}
n_{\rm cr}\equiv \Sigma A_{ji} (j<i) / \Sigma C_{ij} (j\neq i)  
\end{equation}
where $\Sigma C_{ij}$ is 
the collisional (de-)excitation coefficient averaged over 
a Maxwellian-velocity distribution at temperature $T_e$.  
In terms of the dimensionless collision strength $\Omega_{21}$, 
the excitation and de-excitation coefficients are given by  
\begin{equation}
C_{12}={g_2 \over g_1} C_{21} e^{-E_{12}/kT_e},~~ 
C_{21}=8.629\times 10^{-8} {\Omega_{21} \over g_2} 
\left( T_e \over 10^4~{\rm K} \right)^{-1/2}.
\end{equation}
$\Omega_{21}$ weakly depends on temperature.  
For Maxwellian-averaged collision strengths for electron-impact excitation,
we use the results of \cite{tayal2004} and \cite{ramsbottom2007}
for [P II] and [Fe II] lines, respectively.
The critical densities of [P II] and [Fe II] lines 
are 3--$5\times 10^4$ cm$^{-3}$ at $T_e=10,000$~K (Table S2).

The fractional ionization cannot be estimated easily. In collisional ionization
equilibrium, the P II and Fe II fractions peak near unity at 16,000 and
13,000 K, respectively. The ionization fraction curves are similar with 
a slight shift in temperature.
In shocked gas, however, the ionization curves
are affected by time-dependent ionization and photoionization,
so that the fractional ionization could be far from the 
equilibrium value. In the next section, we discuss the structure 
of shocked SN ejecta, and develop a simple model for the abundance analysis.   

\begin{figure}[t!]
\begin{center}
 \includegraphics[width=4.7in]{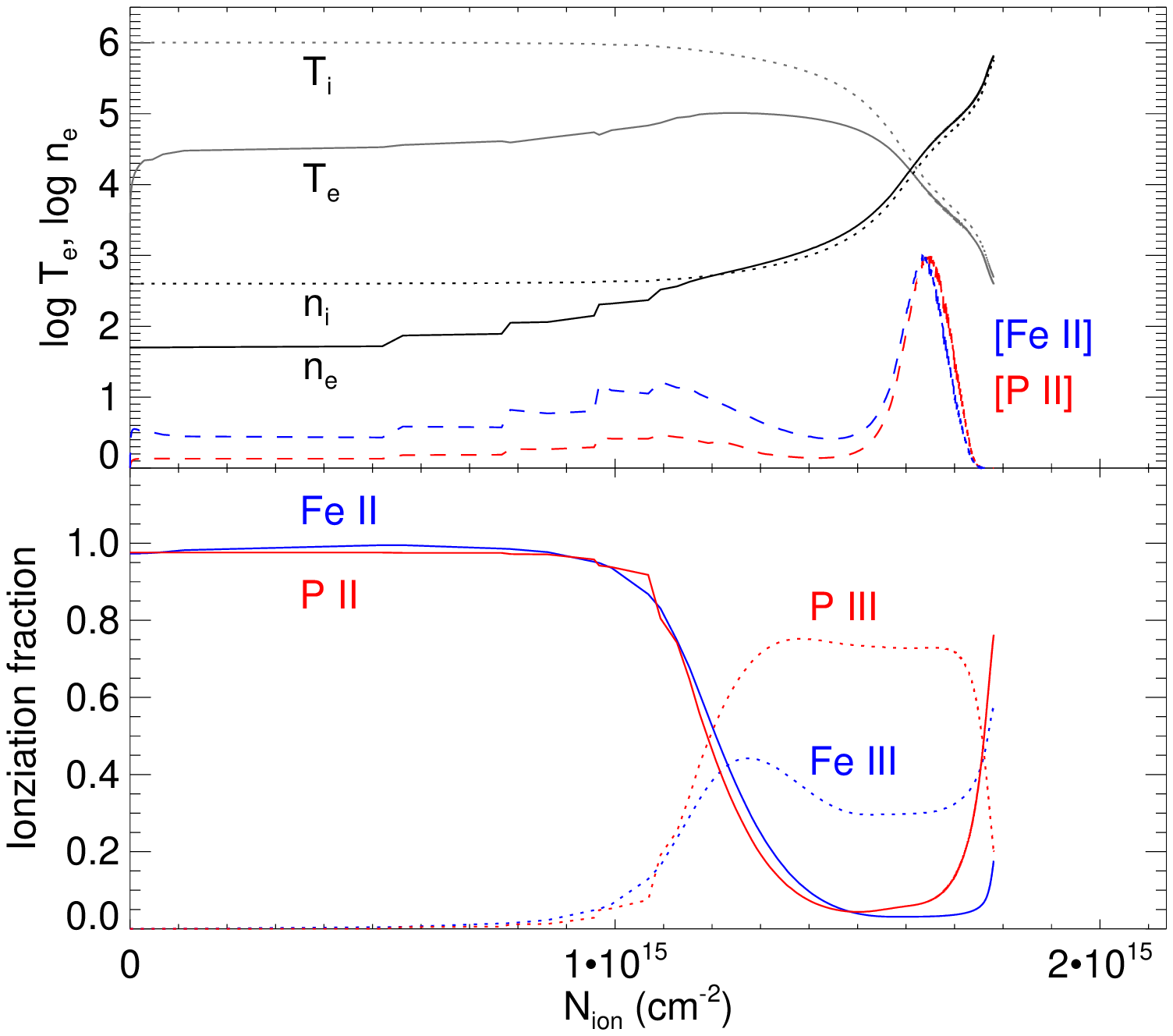}
 \caption{   
(top) Temperature and density profiles of electrons and ions
as a function of swept-up ion-nuclei
column density for a 50~\kms\ shock propagating into an
oxygen-enriched SN ejecta of $n_{\rm ion}=100$~cm$^{-3}$.
[P II] 1.189 $\mu$m and [Fe II] 1.257 $\mu$m
line emissivities per ion (erg s$^{-1}$ sr$^{-1}$)
are overplotted in a linear scale.
(bottom) Ionization fraction profiles of phosphorus and iron.
The calculation is done by using a shock code
developed for SN ejecta \cite{raymond1979,cox1985,blair2000} 
with updated atomic constants for phosphorus. 
}
\label{fig3}
\end{center}
\end{figure}

\noindent
{\bf [P II] and [Fe II] emission from shocked SN ejecta}

The physical structure of shocked SN ejecta has been 
discussed in several previous studies
\cite{itoh1981a, itoh1981b, dopita1984, sutherland1995, 
blair2000, docenko2010}.
A noticeable characteristic of the 
shocks propagating into SN ejecta  
composed of heavy elements is the 
strong cooling that occurs before the ions achieve equilibrium 
ionization states. This is shown in Fig. S3, where 
we plot the temperature and density profiles (top frame)
and ionization fraction profiles of phosphorus and iron (bottom frame)  
of a `typical' dense SN ejecta knot swept up by a reverse shock.
Note that there is an extended ``temperature plateau'' region
where electrons are at temperatures much lower than those of the
ions. For the electron gas in this region,
the heating by Coulomb collisions with ions balances
with the cooling by the collisional excitation and ionization
of the ions. As the electron density increases,
the cooling becomes efficient and the
electron and ion temperatures drop, which happens in
a very narrow region.
In the plateau region,
both phosphorus and iron atoms are mostly in
singly ionized states, ionized by the Galactic
background radiation field, i.e.,
$f_{\rm P II}\approx f_{\rm Fe II}\approx 1$.
In the cooling region, these fractions
decrease as the ions become more highly ionized by electron
collisions. 

The [P II] and [Fe II] emission lines originate from
both the temperature plateau and the cooling regions.
The cooling region is very thin and the fractions of 
the singly ionized ions are low, but these are  
compensated by the increase in electron density, so
that, depending on shock speed, preshock density, and
chemical composition,
the emission from the cooling region can dominate.
In principle, there could be some emission
from the preshock region photoionized by the shock
radiation\cite{dopita1984,sutherland1995,docenko2010},
but in the Cas A knots 
the observed high electron densities 
($3\times 10^3$--$2\times 10^5$ cm$^{-3}$; see below) indicate that
the observed [P II] and [Fe II] emission lines are
mostly from the shocked gas.

In SN ejecta swept-up by a reverse shock,
it is likely that a wide
range of preshock densities and also a wide range of
shock speeds are present\cite{sutherland1995, 
blair2000}. 
We ran a grid of models and have found that,
in order to match the observed [Fe II] line ratios,
the shock speed should be $\simlt 50$~\kms\
and the preshock densities should be $\simgt 100$~cm$^{-3}$.
For such slow shocks in dense SN ejecta, a simple model where
the emitting region is at $T_e=10,000$~K and has an uniform
density (determined from the [Fe II] line ratios)
with $f_{\rm P II}/f_{\rm Fe II}=1$
yields P/Fe abundance ratio accurate to 
within about 30\%. For example, 
the assumed \xpfe\ in the model used in Fig. S3 was 
0.10 while it is 0.090 if we apply the simple model 
to the resulting [Fe II] 1.667/[Fe II] 1.644 and 
[P II] 1.189/[Fe II] 1.257 [Fe II] line ratios.
This is because firstly the upper states of the [P II] 1.189 $\mu$m
and [Fe II] 1.257 $\mu$m lines
have comparable excitation 
energies (12,780 and 11,450 K), so that
the level-population ratio ($f_{\rm ^1D_2, PII}/f_{\rm a^{4}D_{7/2}, Fe II}$)
is almost independent of temperature and is 
mainly a function of electron density $n_e$.
Secondly their  critical densities are comparable
(3--5$\times 10^4$~cm$^{-3}$) and P and Fe have 
comparable ionization energies, i.e., 
10.49 vs. 7.90 eV for I$\rightarrow$II and 19.77 vs. 16.19 eV 
for II$\rightarrow$III, so that 
the two lines are emitted
from nearly the same spatial volume as we see in Fig. S3.
Therefore, we can obtain a reliable estimate of the abundance ratio 
\xpfe\ from a simple model. Note that this 
simple model is also applicable for the circumstellar knots of 
normal chemical composition where the emission is from dense, 
cooling regions.

\begin{figure}[t]
\begin{center}
\includegraphics[width=4.8in]{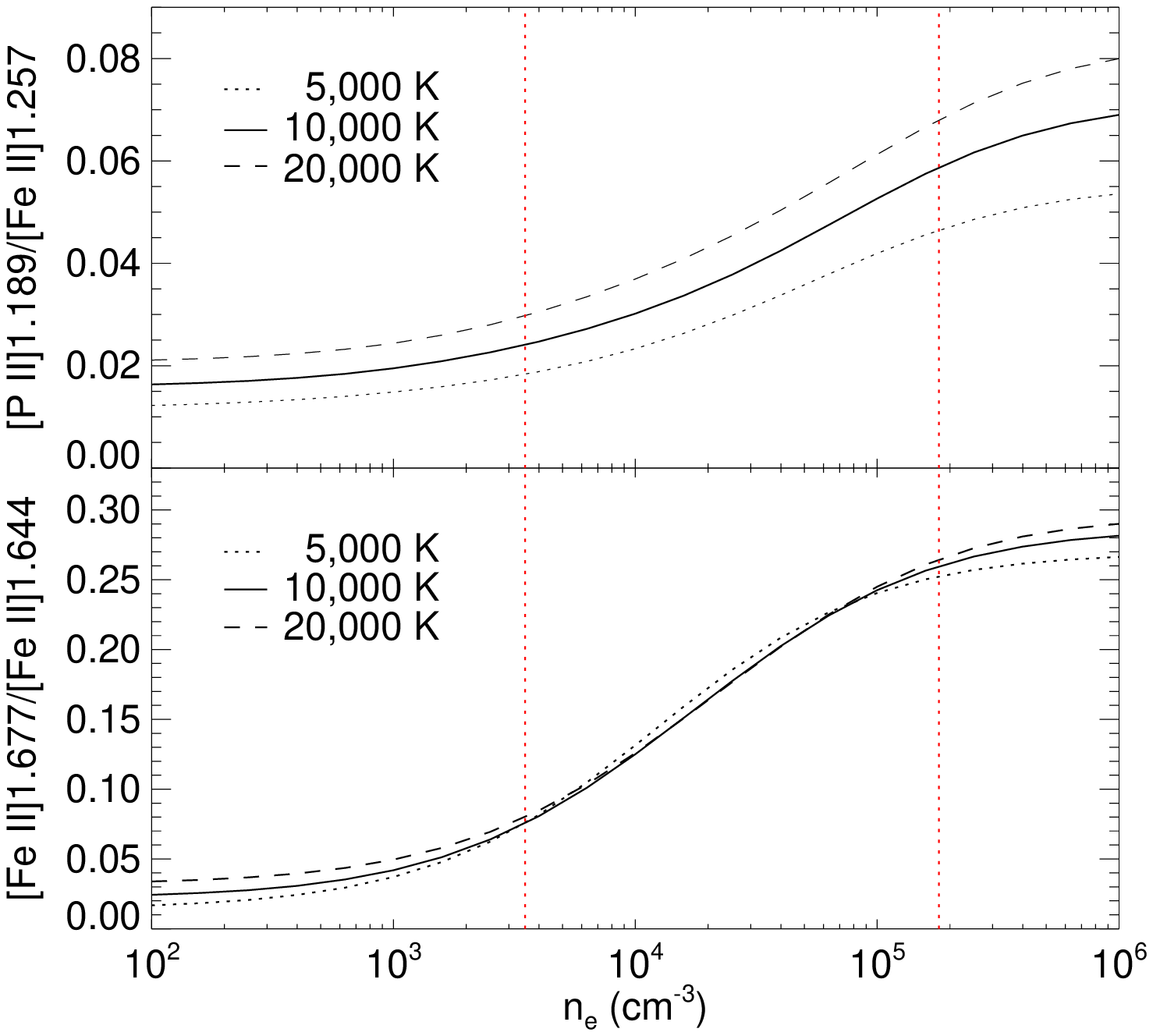}
\caption{
(top) [P II] 1.189/[Fe II] 1.257 line intensity ratio 
for different temperatures. Note that it assumes 
$f_{\rm P II} = f_{\rm Fe II}$ and 
the solar abundance, i.e., $\np/\nfe=8.1\times 10^{-3}$[1].
The range ($\pm 2\sigma$) of the electron densities 
of the Cas A knots are marked by dotted lines.
(bottom) [Fe II] 1.677/[Fe II] 1.644 line intensity ratio as a  
function of $n_e$ for different temperatures.
}
\end{center}
\end{figure}

Fig. S4 (top) shows the [P II] and [Fe II] line ratio 
as a function of electron density at $T_e=10,000$~K. 
We also show the curves for $T_e=5,000$ and 20,000 K 
for comparison.
The temperature dependence is weak as expected.
A great advantage of the broadband NIR spectroscopy is that 
there exist  many [Fe II] lines that can be used to derive electron 
density. One such set is [Fe II] 1.644 and 1.677 $\mu$m lines.
Fig. S4 (bottom) shows that the intensity ratio of the two lines 
is almost independent of temperature. 
Hence, for a given knot, we can first derive 
the electron density from 
${F_{\rm [Fe II]1.677}/F_{\rm [Fe II]1.644}}$, and use this density
to estimate \xpfe.
The mean observed ${F_{\rm [Fe II]1.677}/F_{\rm [Fe II]1.644}}$
ratio is $0.167\pm 0.046$,  
corresponding to $n_e=2.1 \times 10^4$~cm$^{-3}$.
However, the variations among the knots are large, and 
electron densities corresponding to 
$\pm 2\sigma$ deviations from the mean 
($3.5\times 10^3$--$1.8\times 10^5$ cm$^{-3}$) 
are marked by dotted lines in Fig. S4. 
It is worth noting that, 
for the range of the electron densities of the Cas A knots,
the flux ratio can be written as  
\begin{equation}
{F_{\rm [P II]1.189}\over F_{\rm [Fe II]1.257}} 
= (3-7) X({\rm P/Fe}).
\end{equation}
%

\bibliography{scibib}
\bibliographystyle{Science}

{}
\clearpage



\begin{deluxetable}{c c c r r r r r r r r}
\center
\tablecolumns{11}
\tabletypesize{\small}
\setlength{\tabcolsep}{0.05in}
\tablecaption{
{\bf Flux table for the identified near-infrared knots.} 
\hfill\break
}
\tablehead{
\colhead{\bf Slit}      &       \colhead{\bf Knot}      &       \colhead{\bf l}         &       \colhead{\bf v$_{\rm rad}$}             &       \colhead{\bf d}         &
        \multicolumn{6}{c}{\bf F$_{\rm \lambda}$ [10$^{-15}$ erg s$^{-1}$ cm$^{-2}$]}   \\
\cline{6-11}
\colhead{\bf No.}       &       \colhead{\bf No.}               &       \colhead{[$''$]}        &       \colhead{[km/s]}                                &       \colhead{[$''$]}        &
        \colhead{[S II]$_{1.032}$}      &
        \colhead{He I$_{1.082}$}        &
        \colhead{[P II]$_{1.189}$}      &
        \colhead{[Fe II]$_{1.257}$}     &
        \colhead{[Fe II]$_{1.644}$}     &
        \colhead{[Fe II]$_{1.677}$}     \\
}
\startdata

   1   &     1   &    8.75   &      1696   &   118.3   &      5541(32) &   $\leq$46 &     719(14) &     312(13) &     230(4) &      55(3) \\ 
   1   &     2   &    3.25   &       195   &   104.8   &      2226(53) &   $\leq$53 &     253(11) &     136(8) &     100(3) &       8(2) \\ 
   1   &     3   &    5.00   &       -81   &   104.6   &     $\leq$32 &    1430(7) &   $\leq$13 &     290(3) &     213(2) &      38(2) \\ 
   1   &     4   &    5.50   &     -2244   &   103.1   &      1054(6) &   $\leq$8 &      62(4) &      23(3) &      17(3) &   $\leq$2 \\ 
   1   &     5   &    5.75   &     -1100   &   101.8   &      1060(10) &   $\leq$9 &     174(5) &     120(3) &      88(1) &      17(3) \\ 
   1   &     6   &    3.00   &     -1113   &    98.3   &       352(4) &   $\leq$4 &      55(3) &      32(2) &      23(1) &       6(2) \\ 
   1   &     7   &    5.75   &       -87   &    97.3   &     $\leq$164 &   $\leq$79 &   $\leq$36 &     406(15) &     299(6) &      15(4) \\ 
   1   &     8   &    3.00   &     -1252   &    96.6   &       636(4) &   $\leq$5 &     129(3) &      48(2) &      35(1) &       7(3) \\ 
   2   &     1   &    4.50   &       136   &    95.9   &     $\leq$30 &   $\leq$27 &   $\leq$12 &      70(5) &      52(3) &       4(1) \\ 
   2   &     2   &    5.75   &      1176   &    95.6   &      1738(42) &   $\leq$41 &     117(24) &      49(15) &      36(5) &   $\leq$10 \\ 
   2   &     3   &    8.00   &      -702   &    97.6   &      1137(15) &   $\leq$13 &     197(8) &      39(6) &      29(3) &   $\leq$5 \\ 
   2   &     4   &    9.25   &       798   &   106.3   &      1294(26) &   $\leq$41 &      65(11) &     216(9) &     159(4) &      22(2) \\ 
   2   &     5   &    6.50   &      1122   &   106.3   &      1685(105) &   $\leq$61 &     120(23) &     297(27) &     218(8) &   $\leq$10 \\ 
   2   &     6   &    3.75   &      1337   &   108.4   &       983(43) &   $\leq$27 &      29(13) &     106(16) &      78(4) &       5(2) \\ 
   3   &     1   &    6.75   &      -485   &   107.8   &      2282(17) &      93(15) &     269(11) &     252(6) &     185(3) &      33(2) \\ 
   3   &     2   &    5.00   &      -410   &   105.0   &      1259(11) &      60(10) &     116(7) &     153(3) &     113(2) &      21(2) \\ 
   3   &     3   &    7.25   &      -294   &   102.3   &      2960(10) &      68(8) &     273(8) &     460(4) &     339(2) &      60(2) \\ 
   3   &     4   &    6.50   &       194   &    99.6   &       794(13) &   $\leq$40 &      50(7) &     199(5) &     146(3) &      24(1) \\ 
   3   &     5   &    5.75   &       270   &    97.3   &       184(14) &   $\leq$24 &   $\leq$8 &      47(5) &      35(3) &       5(1) \\ 
   4   &     1   &    5.25   &      1002   &    95.3   &     $\leq$396 &   $\leq$144 &   $\leq$76 &     447(56) &     329(9) &   $\leq$7 \\ 
   4   &     2   &    3.75   &      1272   &    98.3   &      1145(41) &     272(31) &   $\leq$20 &     240(13) &     176(3) &      21(2) \\ 
   4   &     3   &    5.75   &      -513   &   100.0   &     $\leq$26 &      57(10) &   $\leq$10 &     144(5) &     106(2) &      21(2) \\ 
   4   &     4   &    6.25   &       859   &   101.5   &      5194(28) &   $\leq$37 &     304(14) &     444(12) &     327(3) &      55(3) \\ 
   4   &     5   &    4.00   &        -8   &   101.5   &     $\leq$16 &   $\leq$14 &   $\leq$7 &      98(3) &      72(2) &      10(1) \\ 
   4   &     6   &    5.75   &       393   &   102.0   &      1626(24) &   $\leq$42 &     134(13) &     449(11) &     330(4) &      49(2) \\ 
   5   &     1   &    5.25   &      1913   &   112.3   &      1465(68) &   $\leq$88 &      85(19) &     302(21) &     222(5) &      33(4) \\ 
   5   &     2   &    5.25   &       983   &   112.2   &      2658(51) &   $\leq$36 &     107(14) &     338(14) &     249(5) &      58(3) \\ 
   5   &     3   &    4.50   &      1350   &   112.1   &     $\leq$91 &   $\leq$25 &   $\leq$20 &     171(13) &     125(4) &      10(2) \\ 
   5   &     4A$^{a}$   &    6.25   &      -118   &   112.2   &      3596(27) &   $\leq$217 &      94(15) &     298(8) &     220(3) &      42(3) \\ 
   5   &     4B$^{a}$   &    6.25   &      -118   &   112.1   &     $\leq$182 &   19435(68) &     107(15) &    1195(8) &     880(3) &     169(3) \\ 
   5   &     5   &    5.25   &      1544   &   111.7   &      1188(31) &   $\leq$42 &   $\leq$27 &     300(13) &     220(3) &      36(2) \\ 
   5   &     6   &    4.75   &      -184   &   111.4   &       854(26) &   $\leq$120 &   $\leq$28 &     165(11) &     121(4) &      25(3) \\ 
   5   &     7   &    3.75   &      -494   &   111.3   &       536(18) &   $\leq$14 &   $\leq$13 &     100(6) &      73(2) &      18(2) \\ 
   5   &     8   &    5.00   &       605   &   110.8   &      1042(52) &   $\leq$76 &     136(22) &     418(17) &     308(6) &      54(3) \\ 
   5   &     9   &    1.75   &      1579   &   110.7   &       882(63) &   $\leq$52 &   $\leq$33 &     101(16) &      74(4) &       7(3) \\ 
   5   &    10   &    9.75   &      -293   &   110.8   &     $\leq$98 &     422(37) &   $\leq$40 &    2564(11) &    1887(4) &     375(4) \\ 
   6   &     1   &    3.50   &      1044   &   114.4   &       717(12) &   $\leq$19 &      17(6) &     102(8) &      75(3) &      18(2) \\ 
   6   &     2   &    3.25   &      1391   &   114.3   &      1150(21) &   $\leq$29 &     136(12) &     215(13) &     158(4) &      21(2) \\ 
   6   &     3   &    4.25   &      1913   &   114.5   &       833(23) &   $\leq$79 &     123(21) &     514(13) &     378(4) &      61(3) \\ 
   6   &     4A$^{a}$   &    2.75   &      -159   &   114.1   &     $\leq$111 &     470(25) &   $\leq$28 &      43(7) &      32(4) &       6(3) \\ 
   6   &     4B$^{a}$   &    2.75   &      -159   &   114.0   &       767(33) &   $\leq$55 &     107(22) &      27(8) &      20(4) &       4(3) \\ 
   6   &     5   &    3.00   &      1303   &   114.1   &       368(34) &   $\leq$17 &   $\leq$16 &     130(9) &      96(3) &      15(3) \\ 
   6   &     6   &    6.25   &      1540   &   113.7   &      2600(65) &     159(25) &     304(20) &     565(20) &     416(5) &      69(3) \\ 
   6   &     7   &    4.50   &       799   &   113.2   &      3858(23) &   $\leq$34 &     247(13) &     318(9) &     234(4) &      48(2) \\ 
   6   &     8   &    2.75   &       560   &   112.9   &       295(20) &   $\leq$27 &   $\leq$35 &     113(7) &      83(3) &      13(2) \\ 
   6   &     9   &    2.75   &       528   &   112.8   &     $\leq$41 &   $\leq$23 &   $\leq$26 &     143(7) &     105(3) &      18(2) \\ 
   6   &    10   &    3.50   &       459   &   112.8   &     $\leq$33 &   $\leq$22 &   $\leq$20 &     134(5) &      99(2) &      17(2) \\ 
   6   &    11   &    6.50   &      -301   &   112.9   &     $\leq$41 &     129(12) &   $\leq$22 &     295(6) &     217(3) &      34(2) \\ 
   6   &    12   &    4.25   &       316   &   112.9   &     $\leq$85 &   $\leq$41 &   $\leq$31 &      67(10) &      50(4) &       7(3) \\ 
   7   &     1   &    3.25   &      1586   &   121.8   &       791(27) &   $\leq$34 &      97(11) &     124(8) &      91(2) &      14(2) \\ 
   7   &     2   &    5.50   &       -73   &   123.5   &     $\leq$23 &     717(15) &      18(6) &     240(4) &     176(3) &      44(1) \\ 
   7   &     3   &    4.00   &        35   &   128.5   &     $\leq$20 &     731(19) &      33(6) &     372(3) &     274(4) &      55(1) \\ 
   7   &     4   &    2.75   &        75   &   129.9   &     $\leq$24 &    1125(20) &      26(6) &     276(4) &     203(5) &      40(1) \\ 
   7   &     5   &    3.25   &       -78   &   131.2   &     $\leq$39 &    4082(19) &      31(9) &     331(5) &     244(4) &      40(2) \\ 
   8   &     1   &    3.25   &       703   &   121.9   &      3294(16) &     105(16) &     229(8) &      96(7) &      71(3) &      10(2) \\ 
   8   &     2   &    6.00   &       818   &   119.2   &      2161(23) &      68(15) &     215(9) &     182(8) &     134(4) &      28(2) \\ 
   8   &     3   &    6.25   &       715   &   114.9   &      2013(16) &      51(12) &     318(9) &     210(7) &     154(4) &      24(2) \\ 
   8   &     4   &    5.25   &       608   &   112.0   &      4789(24) &     266(54) &     595(11) &     235(9) &     173(3) &      26(2) \\ 
   8   &     5   &    4.50   &       613   &   109.4   &      4484(66) &   $\leq$60 &     567(24) &     145(20) &     107(6) &      11(3) \\ 
   8   &     6   &    4.00   &       473   &   108.6   &     12167(143) &   $\leq$140 &     855(44) &     192(28) &     141(6) &      20(3) \\ 
   8   &     7   &    4.75   &       640   &   106.5   &      4108(67) &   $\leq$45 &     681(18) &     255(16) &     188(5) &      28(3) \\ 
   8   &     8   &    4.00   &       122   &   106.4   &       331(15) &   $\leq$47 &     113(8) &     103(6) &      76(4) &      16(2) \\ 
   8   &     9   &    4.00   &       359   &   105.0   &       461(19) &   $\leq$31 &     210(11) &     158(8) &     117(3) &      22(2)  
\enddata
\tablecomments{
Columns 1--5 present  
slit number, knot number,  angular size along the slit ($l$), 
radial velocity ($v_{\rm rad}$), and 
distance ($d$) from the explosion center 
($\alpha = 23^{\rm h}23^{\rm m}27^{\rm s}.77 \pm 0^{\rm s}.05$,
$\delta = 58^{\circ} 48' 49''.4 \pm 0''.4$ [J2000])[28].
The slit number starts from the top and increases in a 
counterclockwise direction (Fig. 1 in Report). 
Note that Slits 5 and 6, represented by relatively short white bars,  
are located side by side with the former slightly inside.
Columns 6--11 present extinction-corrected fluxes of [S II] 1.032, 
He I 1.082, [P II] 1.189, 
[Fe II] 1.257, [Fe II] 1.644, and [Fe II] 1.677 $\mu$m lines. 
The intensity ratio of the [Fe II] 1.257 and [Fe II] 1.644 $\mu$m lines
is constant (1.36) because they share the same upper level (see text). 
The number in bracket denotes a 1$\sigma$ error of each flux from 
the Gaussian fit, whereas the upper limits are $3\sigma$ limits.
}
\tablenotetext{a}{These knots in Slits 5 and 6 have been identified as a single knot 
by {\it Clumpfind} respectively, but 
a detailed inspection revealed that each of them are composed of two (A and B) 
components almost coincident both in space and velocity.}
\end{deluxetable}
\clearpage

\begin{deluxetable}{ccccccc}
\tablecaption{
{\bf Parameters of near-infrared [P II] and [Fe II] lines used in the analysis.} 
\hfill\break
}
;\tablewidth{0pc}
\tablecolumns{7}
\tablehead{
\colhead {} &  \multicolumn{2}{c}{transition levels} & \colhead{$\lambda$}
& \colhead {$A_{21}$} & \colhead {$n_{\rm cr}$} & \colhead {$T_{\rm ex}$} \\
\colhead{ion} & \colhead {lower} & \colhead{upper} & \colhead{($\mu$m)}
& \colhead{(s$^{-1}$)} &  \colhead{(cm$^{-3}$)}  &  \colhead{(K)}
}

\startdata

P II  &  $^{3}P_{2}$  &  $^{1}D_{2}$  &   1.18861  &  $1.43\times 10^{-2}$ & $5.3\times 10^4$ & 12,780 \\
      &  $^{3}P_{1}$  &  $^{1}D_{2}$  &   1.14713  &  $5.24\times 10^{-3}$ & $5.3\times 10^4$ & 12,780 \\
      &  $^{3}P_{0}$  &  $^{1}D_{2}$  &   1.12583  &  $1.85\times 10^{-6}$ & $5.3\times 10^4$ & 12,780 \\
\\
Fe II &  $a^{6}D_{9/2}$ &  $a^{4}D_{7/2}$ &   1.25702  &    $5.27\times 10^{-3}$ &   $3.5\times 10^4$ & 11,446 \\       
      &  $a^{4}F_{9/2}$ &  $a^{4}D_{7/2}$ &   1.64400  &    $5.07\times 10^{-3}$ &   $3.5\times 10^4$ & 11,446 \\ 
      &  $a^{4}F_{7/2}$ &  $a^{4}D_{5/2}$ &   1.67734  &    $2.11\times 10^{-3}$ &   $3.1\times 10^4$ & 12,074 \\

\enddata

\tablecomments{
Columns present ion, lower and upper transition levels, 
wavelength of the lines ($\lambda$), Einstein $A$ values, 
critical density ($n_{\rm cr}$) at $T_e=10,000$~K (see eq. [3] in the text),
and excitation energy ($T_{\rm ex}$) of the upper level.
The Einstein $A$ values are from 
\cite{mendoza1982} and
\cite{deb2011} for [P II] and [Fe II], respectively. 
}

%
\end{deluxetable}
\clearpage
